\documentclass[authoryear,preprint,review,12pt]{elsarticle}
\usepackage{rotating}
\usepackage{array}
\usepackage{url}
\usepackage{hyperref}
\usepackage{amssymb}
\usepackage{amsmath}

\journal{Weather and Climate Extremes}

\begin{document}
\begin{frontmatter}
\title{Statistics of Temperature Extremes and Implications for Electrical Power Infrastructure in Continental France}

\author[1]{Peter Werner\corref{cor1}} 
\ead{peter.werner@lmd.ipsl.fr}
\author[2]{Bastien Cozian}
\author[3]{Yoann Robin}
\author[2,4]{Laurent Dubus}
\author[1]{Freddy Bouchet}

\cortext[cor1]{Corresponding author}
\affiliation[1]{organization={Laboratoire de Météorologie Dynamique},
addressline={École Normale Supérieure},
postcode={75231},
city={Paris},
country={France}}
\affiliation[2]{organization={Réseau de Transport d'Electricité},
addressline={7C PLace du Dôme},
postcode={92073},
city={Paris La Défense},
country={France}}
\affiliation[3]{organization={Laboratoire des Sciences du Climat et de l'Environnement},
addressline={CEA Saclay - L'Orme des Merisiers},
postcode={91191},
city={Gif-sur-Yvette},
country={France}}
\affiliation[4]{organization={World Energy \& Meteorology Council},
addressline={The Enterprise Centre, University Drive, University of East Anglia},
postcode={NR4 7TJ},
city={Norwich},
country={United Kingdom}}

\begin{abstract}
In this article, we present extreme value statistics of temperature extremes over
continental France with a focus on their implications for
electrical power infrastructure. These are obtained by fitting a non-stationary
generalised extreme value distribution using a Bayesian setup, which also 
provides errors or uncertainty values for all estimates. Within this method, a
combination of simulated data from a collection of 28 CMIP6 models and measured
records from the E-OBS dataset is used. The investigated climate scenarios are
SSP2-4.5, SSP3-7.0 and SSP5-8.5, considering both historical and future climates
spanning the years from 1850 to 2099. The method provides full spatial
resolution on a 0.25 degree grid, allowing to asses extreme temperatures at arbitrary
locations. Leveraging this particular aspect, various maps revealing the spatial
structure of annual maximum high temperature extremes over continental France
for the median and the statistical upper bound are shown together with summary
information for the administrative regions in France towards the end of the
century. It is statistically possible that the south west region of Occitanie could reach
temperatures of up to 57$^{\circ}$C under a high emission scenario by 2080,
which is the highest compared to all other regions in continental France. We also perform a
comparison to the reference climate adaptation trajectory for France
(\textit{TRACC - Trajectoire de réchauffement de Référence pour l'Adaptation au
Changement Climatique}), showing that it potentially underestimates
the stated maximum temperatures which could be surpassed by up to +8$^{\circ}$C.
Furthermore, five reference electrical power infrastructure locations are
investigated on how they are potentially affected by temperature extremes with
the quantified intensities.
\end{abstract}







\end{frontmatter}


\section{Introduction}
\cite{forzieri_2018} estimated that the expected annual damage to the
energy sector in Europe could reach 8.2\texteuro\ billion [5.0-10.7\texteuro\ billion] by 2080.
The largest contribution to these damages was attributed to the sensitivity of energy
production to droughts (62\%) and heatwaves (27\%).
For France, a concrete example from the past is the 2003 heat wave that was estimated to have caused
costs of about 330\texteuro\ million for Électricité de France (EDF), the main electricity company of the country \citep{dubus_2010}.
More recent reviews of past global events \citep{hawker_2024, anel_2024} show not only the negative effects of extreme
temperatures on energy infrastructure, but also highlight the importance of long term planning and preparedness for mitigation.
In this context \cite{hawker_2024} pointed out that because of a warming climate, as a result
of greenhouse gas emissions, knowledge from past events are not reflective of future situations.
Specifically, frequency and intensity of extreme high temperature events will increase, against which existing
and future power infrastructure must be resilient to ensure a reliable energy supply.

The exact physical causes and mechanisms by which extreme temperatures affect power infrastructure are diverse
and include composite events, e.g., heat waves that may lead to wildfires, that can amplify the possibility of damage.
Generally, there are three main aspects:
a) electric energy production, b) electric energy consumption and c) the effects on the physical power infrastructure assets (power lines, power conversion stations...)
We state some example effects caused by extreme temperatures for each aspect in the following to give a general idea, while
a more detailed overview of extreme weather events and their impact on power systems can be found in \citet{bromberger_2026}:
a) Electric energy production: 
During heat waves, thermal power plants lose efficiency, i.e., they need more fuel for the same amount of power output,
since a higher air temperature generally reduces the cooling capabilities from the surroundings \citep{huguet_2008}.
In addition, heat waves are often accompanied by droughts, and the consequentially reduced availability of water at river sites
poses an additional challenge to facilitate cooling for power plants at these locations (which is the case for most nuclear power plants in France).
With decreased river flow, run-of-river hydropower generation is decreased, while at the same time, reservoirs can experience lower filling rate and increased evaporation.
b) Electric energy consumption:
cold waves can lead to increased consumption of electricity for heating, while
heat waves are usually accompanied by higher power consumption due to increased demand for refrigeration and air conditioning.
c) Effects on infrastructure buildings/constructions:
low temperatures in conjunction with precipitation can lead to the accumulation of ice and snow on overhead transmission lines \citep{faggian_2024}.
The resulting increase in weight and wind attack surface, e.g., through icicles, increases mechanical stress on the power lines and
poles, potentially causing snapping or collapse, respectively.
Similarly, tree branches breaking away due to the accumulation of snow pose another risk for failure if they hit overhead power lines.
Ground-ice build-up can affect run-of-river power plants, and sufficiently low temperatures can crack isolators or ropes \citep{rothstein_2010}.
High temperatures also impact transmission lines. At a given current, an increase in ambient temperature causes the cables to expand,
potentially exceeding the minimum safety clearance from the ground. This is undesirable, and the current flowing through the cables must
therefore be reduced to avoid risks to people and the environment.
Extremely high temperatures can also cause damage to the cables, as well as to components of power transformer stations.
Similarly, extremely high temperatures could damage underground lines and necessitate limiting energy transit to avoid
overheating the surrounding soil or causing material damage.
\citet{sergio_network_planning_2025} found that for future climates and depending on cable type, a reduction in thermal
rating ranging from 3.8\% to 6\% is possible for underground lines by 2070 in France.
\citet{sergio_2025} estimated that on average by 2070 over Europe, the thermal rating under a RCP8.5 scenario will decrease
by 1.53\% for overhead lines, by 2.1\% for power transformers and by 0.2\% for underground lines.
During the aforementioned 2003 heat wave, there were also reports of underground medium- to low-voltage cable failures in the Paris area \citep{rothstein_2010}.

Due to this susceptibility of the energy sector to weather and the high costs involved, risk management is imperative \citep{troccoli_2010}
and knowledge regarding current and future statistics on extremes is an important component of it.
An approach of choice for this is extreme value theory \citep{bousquet_2021} and specifically the analysis by means of a \emph{Generalized Extreme Value Distributions} (GEV) \citep{robin_2020, robin_2026}, which has been a frequent tool in climate science to quantify the statistics of extreme weather events.
Naturally, it was also used to gain insight on how extreme events might affect power infrastructure specifically.
For example, \citet{huguet_2008} investigated upstream river temperature for nuclear power plants in France and
\citet{parey_2008} obtained estimates of extreme temperatures to aid planning of power plant constructions.
GEV distributions were also used to asses ice and wind hazards on overhead high voltage power lines \citep{davalos_2023, faggian_2024}.
\citet{milojevic_2023} studied the statistics of extreme precipitation and reservoir inflow events affecting hydropower production.
Beyond the assessment of risk from extreme weather events, GEV distributions were also used in the
description of photovoltaic cell temperatures in tropical regions \citep{yaacob_2014},
and the analysis of maximum available wave power \citep{sardana_2024}.

The aim of this paper is to present analysis results regarding the GEV distribution of temperature extremes over continental France and to put
them into context with power infrastructure, demonstrating the effectiveness of the Bayesian approach of~\citet{robin_2026}
for climate risk assessment.
For this purpose, we discuss several specific infrastructure locations in detail as reference cases on how extreme high temperatures might affect them in the future.
A current shortcoming in studies using GEV distributions to assess power infrastructure risk is the lack of combined spatial and time resolution.
This means that it is desirable to obtain extreme value statistics ideally for time horizons towards the end of the century, and at arbitrary
locations simultaneously, instead of only focusing 
on representative or close-by measurement stations, which was predominantly done in similar studies in the past \citep{parey_2019, hamdi_2018} for France.
In \citep{parey_2019}, for example, projections using data from 13 CMIP5 climate model simulations were obtained up to the year 2100, but only
at nine individual locations across France.
Another methodological issue is that estimates from climate simulations and past observations are mostly treated independently (again see \citep{parey_2019}).
The method employed in this paper alleviates both of these deficiencies by combining data from an ensemble of climate models and observational records through
a Bayesian approach. In this setup, an initial a priori estimate from climate simulation data is constrained by the measurement data to 
obtain an a posterior estimate with overall lower error.
We consider several climate scenarios in the following, properly incorporating a non-stationary climate, over a period towards the end of the twenty-first 
century.

The structure of the paper is as follows:
Section \ref{sec:methodology} briefly introduces the statistical model of the non-stationary GEV distribution
and the steps of the employed Bayesian parameter estimation method together with the utilised datasets.
We also present the summary statistics to characterise the estimated distributions and an explanatory example.
Results are discussed in section \ref{sec:results}, which encompasses maps of mid- to end-century values of GEV summary statistics
for continental France and time-resolved local statistics at selected power infrastructure sites.
A comparison to \citep{tracc_report_part2}, the \textit{Trajectoire de réchauffement de Référence pour 
l'Adaptation au Changement Climatique} (TRACC), which is the official planning 
and adaptation reference with regard to climate change for France, is done.
The conclusion summarizes the results and draws perspectives and recommendations for future work.

\section{Methodology}
\label{sec:methodology}
The detailed description of the Bayesian GEV parameter estimation can be found in~\citet{robin_2020,robin_2026}.
Here we present only the basic aspects and notations.

\subsection{Statistical model}
The quantities of interest are annual maxima of $d$ days-long averaged daily 
maximum 2 m surface air temperature values, denoted by $T_t$ hereafter.
Their statistics usually depend on time $t$, due to a changing climate, and 
are assumed to follow an equally time dependent, i.e., non-stationary,
\emph{Generalized Extreme Value Distribution} (GEV).
We describe the non-stationary GEV distribution through the statistical model
\begin{equation}\label{eq:statistical:model}
	\left\lbrace
	\begin{array}{cl}
        T_t &\sim \textrm{GEV}(\mu_t, \sigma_t, \xi_t) \\
       \mu_t &= \mu_0 + \mu_1 X_t \\
       \log(\sigma_t) &= \sigma_0 + \sigma_1 X_t \\
       \xi_t &\equiv \xi_0 \\
       X_t &= X^0 + X^N_t + X^A_t
	\end{array}
	\right. , 
\end{equation}
where $\textrm{GEV}(\mu_t, \sigma_t, \xi_t)$ is the probability density function of the stationary GEV distribution
(see Appendix \ref{sec:gev:distribution} for formula) but with time dependent parameters
for location $\mu_t \in \mathbb{R}$, scale $\sigma_t > 0$, and shape $\xi_t \in \mathbb{R}$.
The logarithm around $\sigma_t$ ensures its positivity.
The time dependency of the GEV parameters is realized through coupling 
functions in which a linear relationship to a covariate $X_t$ is assumed that consist of
a constant contribution $X^0$ and contributions from natural forcings $X^N_t$ and anthropogenic forcings $X^A_t$.
The last term $X^A_t$ is given by a smoothed curve, realized through spline functions. The annual mean temperature over Europe
serves as an approximation of the covariate $X_t$.
For notational purposes, we collect all parameters here into one symbol
$\boldsymbol{\theta}  := (X^0, X^N_t, X^A_t, \mu_0, \mu_1, \sigma_0, \sigma_1, \xi_0)$ in the following.

\subsection{Bayesian parameter estimation \& datasets}
\begin{sidewaysfigure*}[t]
    \centering
	\includegraphics[width=\linewidth]{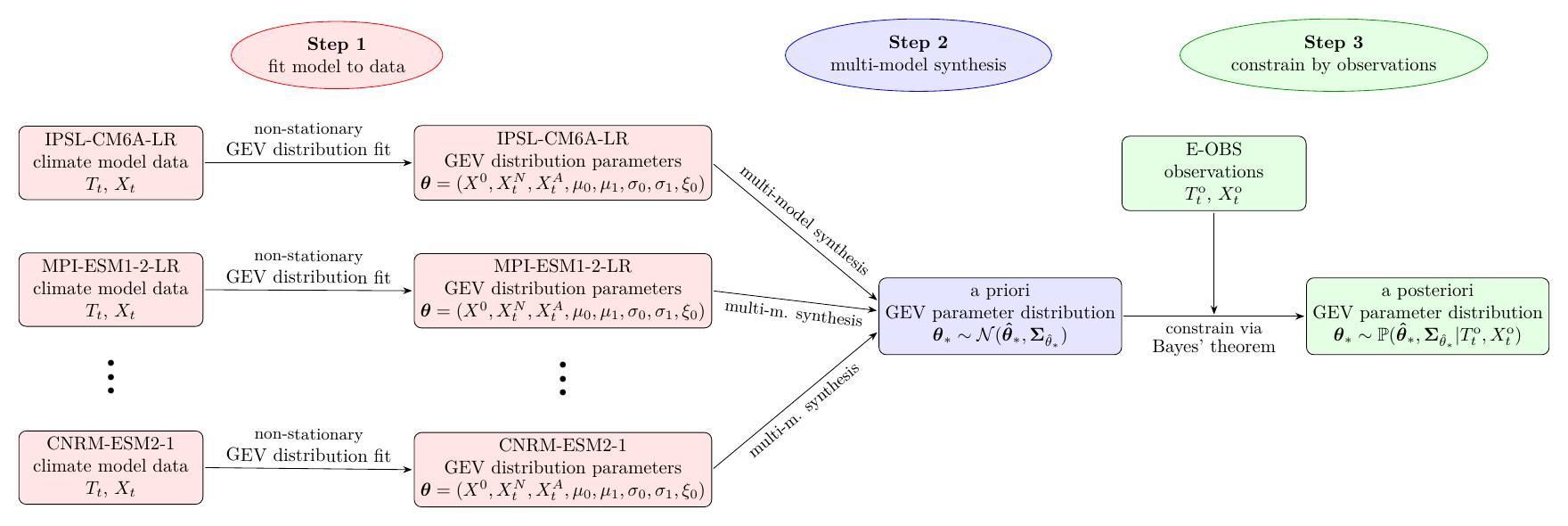}
	\caption{\label{fig:method:description}
    Graphical representation of the Bayesian data analysis procedure. The method consist of 
    three steps, where nodes that belong to the same step have the same colour.
    The nodes themselves represent data while arrows indicate a form of data processing.
    For Step 1, only three of all 28 climate models, which are listed in Table \ref{tab:cmip6:models}, are explicitly shown.}
\end{sidewaysfigure*}

The final goal of the numerical analysis is to obtain estimates, including an 
appropriate error or uncertainty range, for the non-stationary GEV distribution parameters.
Special about the outlined approach is that it is the only known procedure to combine measured data and model data
from climate simulations in a statistically consistent way.
We use the software package ANKIALE \citep{robin_2026, robin_github_2025} in version 1.0.3 to facilitate the individual steps of the analysis.
Spanning the period from 1850 to 2099, we investigate simulation data from the historical period and for the future
\emph{Shared Socioeconomic Pathways} \citep{van_vuuren_2017} SSP2-4.5, SSP3-7.0 and SSP5-8.5.
Figure~\ref{fig:method:description} shows a graphical representation of the procedure, which consists of three steps.
In the first step (coloured in red), the statistical model in Eq. \eqref{eq:statistical:model} is fitted to temperature data
for $T_t$ and $X_t$ from 28 CMIP6 climate simulations that are listed in Table \ref{tab:cmip6:models}, amounting to 28 individual
parameter estimates $\boldsymbol{\theta}$.
During the second step (blue), an a priori distribution of the non-stationary GEV parameters
is determined from multi-model synthesis, following a "models are statistically indistinguishable from the truth" paradigm \citep{ribes_2017}.
The a priori distribution is assumed to be Gaussian $\mathcal{N}(\boldsymbol{\hat{\theta}}_*, \boldsymbol{\Sigma}_{\hat{\theta}_*})$
with mean $\boldsymbol{\hat{\theta}}_*$ and covariance $\boldsymbol{\Sigma}_{\hat{\theta}_*}$.
The third step (green) is constraining the a priori parameter estimate with observational data.
For observations, we use the E-OBS dataset in version 31.0e with 0.25 degree grid
resolution \citep{cornes_2018, eobs_v31e} for both the covariate $X^{\mathrm{o}}$, i.e, the mean annual temperature over Europe,
and the annual temperature maxima $T^{\mathrm{o}}$.
The ANKIALE software interpolates values on the CMIP6 model grids to the E-OBS 0.25 degree grid.
The constraint itself is done via Bayes' theorem \citep{sivia_2006} and consists of a combination of the Gaussian
conditioning theorem and a Markov Chain Monte-Carlo maximisation scheme.
This yields the a posteriori parameter distribution
$\mathbb{P}(\boldsymbol{\hat{\theta}}_*, \boldsymbol{\Sigma}_{\hat{\theta}_*} | T^{\mathrm{o}}_t, X^{\mathrm{o}}_t)$.
For each year and grid cell, we draw 1000 statistically independent samples $\boldsymbol{\theta}_*$
from the a posteriori distribution.
These samples are used to determine estimates, including uncertainties for the summary statistical
quantities outlined in the following subsection.

\subsection{GEV summary statistics}
In order to assess the estimated GEV distributions in relation to electrical power infrastructure,
we evaluate several summary statistics.
First, typical extreme temperatures are well described by the median, which is given by the $p=0.5$ quantile.
The quantile formula for the GEV distribution is:
\begin{equation}\label{eq:gev:quantile}
	\mathrm{Q}_t(p|\boldsymbol{\theta}_*) = \left\lbrace \begin{array}{cl}
		\mu_t - \sigma_t \ln(-\ln(p)) & \text{for } \xi_t = 0\\
		\mu_t + \frac{\sigma_t}{\xi_t}\left( (-\ln(p))^{-\xi_t} - 1\right) 
		& \text{for } \xi_t \neq 0
	\end{array} \right. .
\end{equation}
Second, the upper bound
\begin{equation}\label{eq:gev:upper:bound}
    B_t = \mu_t - \sigma_t / \xi_t
\end{equation}
itself gives indications on worst-case events since it describes the maximal possible temperature.
Lastly, the difference between the upper bound and the median, denoted by
\begin{equation}\label{eq:gev:delta}
	\Delta_t(\boldsymbol{\theta}_*) := B_t  - \mathrm{Q}_t(0.5|\boldsymbol{\theta}_*),
\end{equation}
is a measure for the range of possible extreme temperatures that 
infrastructure could be exposed to in the future, i.e., $\Delta_t$ can be seen as 
a sort of maximum range of high operation temperatures.

In particular, the posterior samples allow for an uncertainty estimate of the
summary statistics in Eq. \eqref{eq:gev:quantile}, Eq. \eqref{eq:gev:upper:bound}
and Eq. \eqref{eq:gev:delta}, which is given in the 
course of this paper by the interquartile range, i.e., the difference between 
the 0.75 and 0.25 quantiles, over all samples.
It will be referred to as uncertainty or error in the following.
To be specific, there are two types of distinct quantiles here: 
First, the GEV distribution quantiles in Eq. \eqref{eq:gev:quantile}, which 
characterize the distribution for one particular sample with values $\boldsymbol{\theta}_*$.
And second, the posterior distribution quantiles, which serve to characterise the uncertainty of the estimate 
on the non-stationary GEV distribution parameters themselves.

\subsection{Explanatory example}
\begin{figure}[t]
    \centering
	\includegraphics[width=\linewidth]{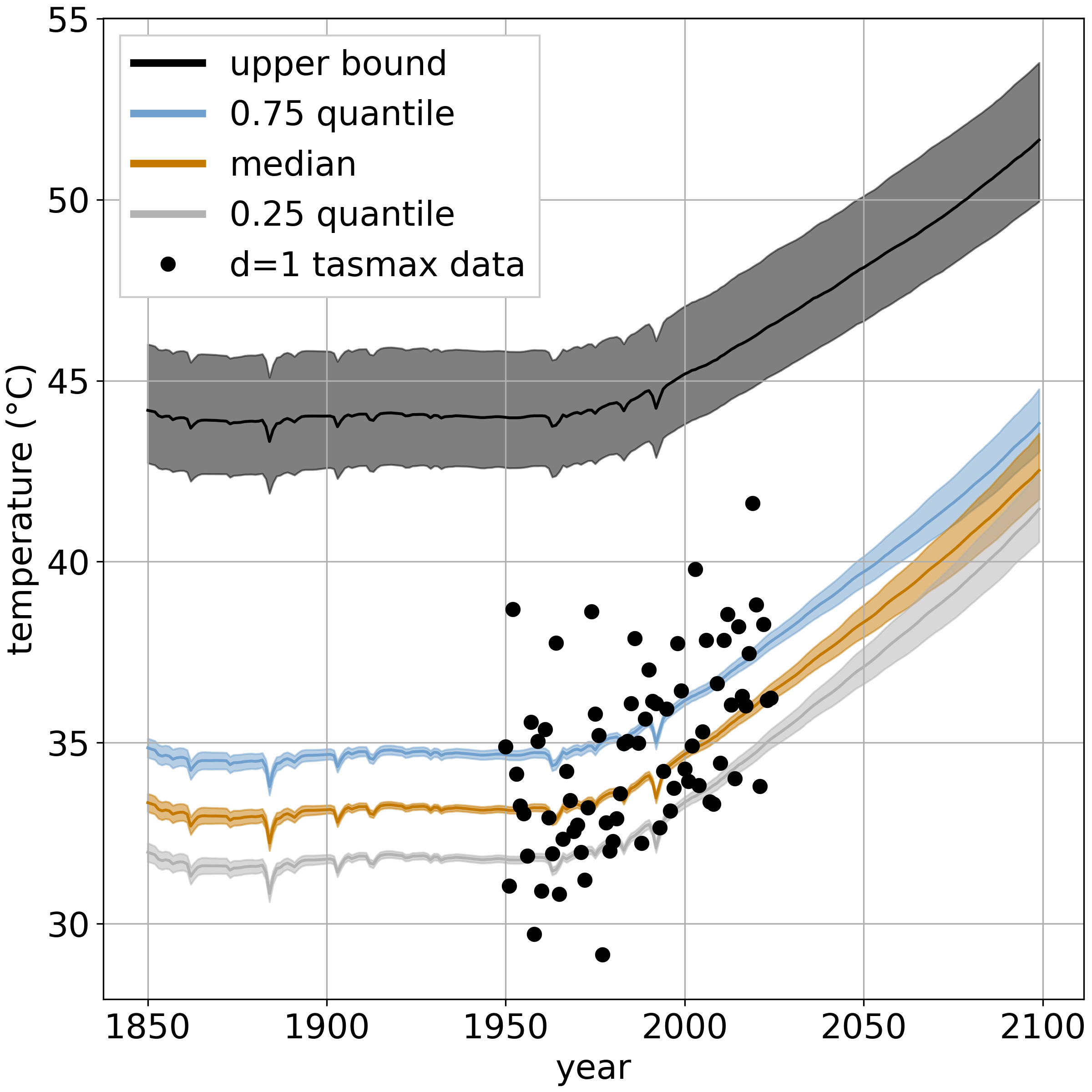} 
	\caption{\label{fig:explanatory:plot}
	Estimated non-stationary GEV distribution quantiles and upper bound (see  
	Eq. \eqref{eq:gev:quantile} and Eq. \eqref{eq:gev:upper:bound}) for the 
	annual maximum of daily maximum surface air temperatures under the SSP3-7.0 scenario at 
	the Marmagne power station (see Table \ref{tab:infrastructure:locations}).
	The shaded regions represent the error range.
	The observational data is shown as black points.
    A quarter of the data points is expected to be above the blue 0.75 quantile line,
    Half of the data points are expected to be above or below the orange median curve.
    Another quarter of the data points is expected to be below the grey 0.25 quantile line.
    None of the data points is above the black upper bound curve.}
\end{figure}

To elucidate the method and to aid comprehension, we depict an exemplary result at the grid point around the Marmange power station
(coordinates are in Table \ref{tab:infrastructure:locations}) for the SSP3-7.0 scenario in Fig. \ref{fig:explanatory:plot}.
Such estimates of the GEV distribution are calculated by the ANKIALE software at every grid point, where the grid matches that of the E-OBS dataset used.
Shown is the non-stationary GEV distribution, in terms of its 0.75, 0.5 and 0.25 quantiles from Eq. 
\eqref{eq:gev:quantile} and upper bound (Eq. \eqref{eq:gev:upper:bound}), as a function of time.
The observational data are depicted as black points, which were used to constrain the a priori GEV parameter distribution.
The expectation is that 25\% of the data points should be above the 0.75 quantile curve or below the 0.25 quantile curve.
Half of the data points are expected to be in between the 0.75 and 0.25 quantile curves.
The median, i.e., the 0.5 quantile, divides the number of data points on average into 50\% above and
50\% below its curve.
None of the data points can surpass the upper bound, as it is the definition range of the distribution.
For the example in Fig. \ref{fig:explanatory:plot}, the observational data points correspond reasonably well
within statistical errors with these expectations towards the curves for quantiles and upper bound.

Usually, values at a particular quantile can also be associated with a corresponding return time.
For the present case of annual values, the 0.75 quantile marks events with equal or higher temperature values that are expected to occur on average every four years.
However, this direct identification with return times is no longer appropriate in a non-stationary context \citep{hamdi_2018}.
In a warming climate, as in Fig. \ref{fig:explanatory:plot}, this simply means that a certain quantile value,
increases over time and consequently happens more frequently, which would not be the case in a non-changing climate.
Therefore, an evaluation conditioned to a particular year of interest is more appropriate.
This is also why return levels, i.e., the quantiles corresponding to certain return periods, are omitted here.
Even though return levels are commonly used in risk assessment, and alternative definitions adapted to the non-stationary case exist \citep{parey_2007}, 
here the upper bound is used to quantify the worst-case temperatures, since it constitutes a hard limit 
on the temperature exposure and associated risks.

Figure~\ref{fig:explanatory:plot} also provides an overview of the general behavior and trends in the
distribution of annual extreme temperatures.
From 1850 to roughly 2000 the distribution is mostly stationary, i.e., its quantiles do not change over time, with
50\% of events with temperatures higher than around 33$^{\circ}$C.
Towards the end of the century, under the SSP3-7.0 scenario, it is expected that
50\% of events will be hotter than about 42$^{\circ}$C.
Naturally, the uncertainty of these estimates, indicated by the shaded regions, 
generally increases as the prediction becomes more distant in the future.
In particular, the upper bound, i.e. the statistically maximum possible 
temperature is usually difficult to estimate and has overall rather
large error bars in comparison to, for example, the median.
However, the plot indicates that the upper bound in the
past century is around 44 $^{\circ}$C, while at the
end of the twenty-first century, the estimate is around 52$^{\circ}$C.
This means that certain temperature extremes that were considered impossible
in the past will be possible in the future.
It also underlines the fact that extreme high temperature events observed in the past
might not be representative of future events in terms of intensity, which may be an important 
consideration during the planning of future power infrastructure.

\section{Results}
\label{sec:results}
In the following subsections, the spatial structure over continental France is investigated 
and at certain infrastructure locations, the time-resolved properties of the non-stationary GEV distribution are presented.
A comparison to the TRACC is also provided.
We put the obtained statistical information about temperature extremes into 
context with temperature-related criteria for electrical power infrastructure.

\subsection{Regional differences of extreme temperatures in continental France}
\begin{figure*}[t]
    \centering
	\includegraphics[width=\linewidth]{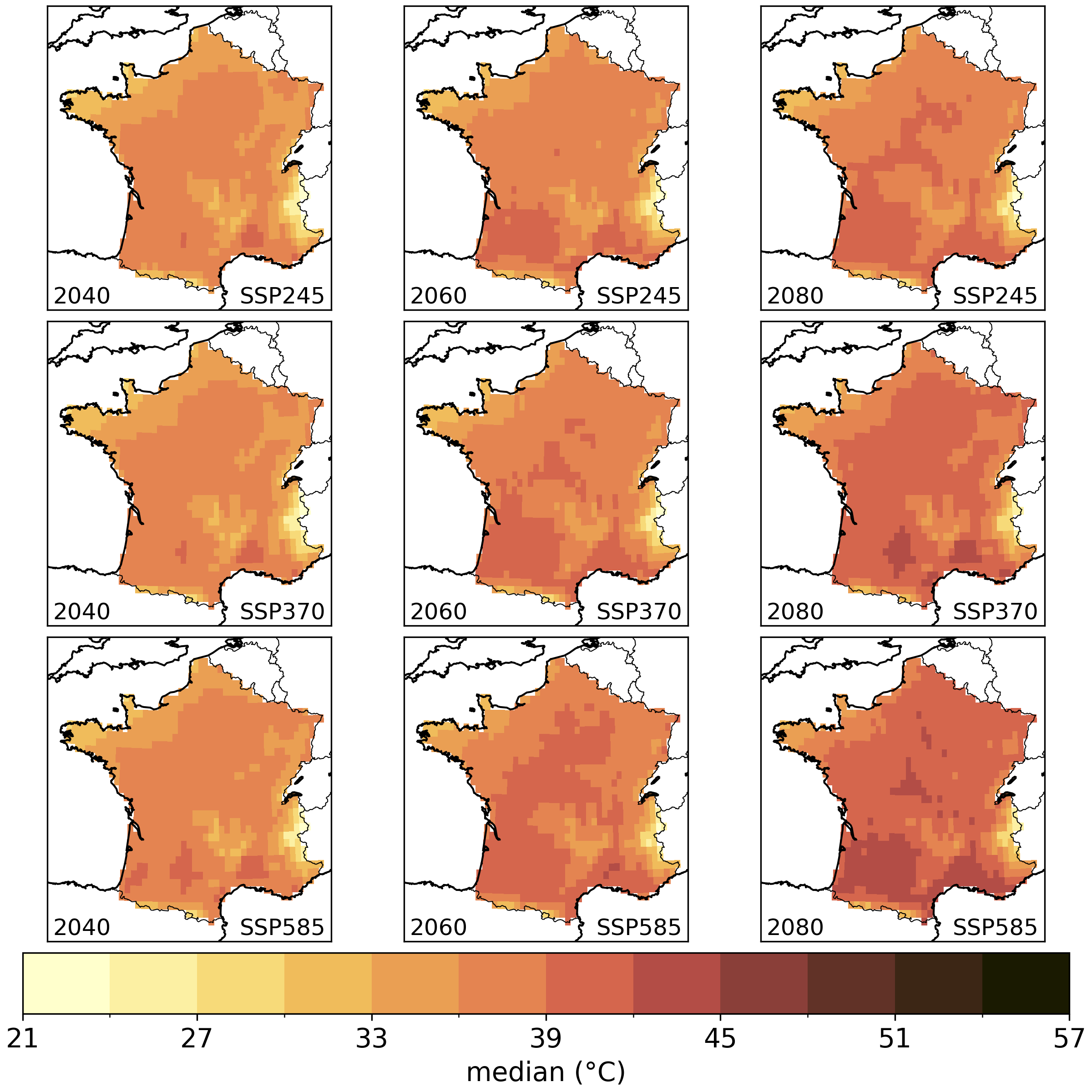}
	\caption{\label{fig:map:median}
	Maps of continental France for the median $\mathrm{Q}_t(0.5|\boldsymbol{\theta}_*)$ 
	(see Eq. \eqref{eq:gev:quantile}) of the GEV distribution for annual 
	maxima of daily maximum surface air temperature values in the years 2040, 
	2060 and 2080 (columns) and for the climate scenarios SSP2-4.5, SSP3-7.0 and SSP5-8.5 (rows).
	Over all depicted maps, the values of $\mathrm{Q}_t(0.5|\boldsymbol{\theta}_*)$ range 
	from 22.4 $^{\circ}$C to 44.7 $^{\circ}$C. The associated uncertainties (not shown here)
    range from 0.4 $^{\circ}$C to 2.4 $^{\circ}$C.}
\end{figure*}

\begin{figure*}[t]
    \centering
	\includegraphics[width=\linewidth]{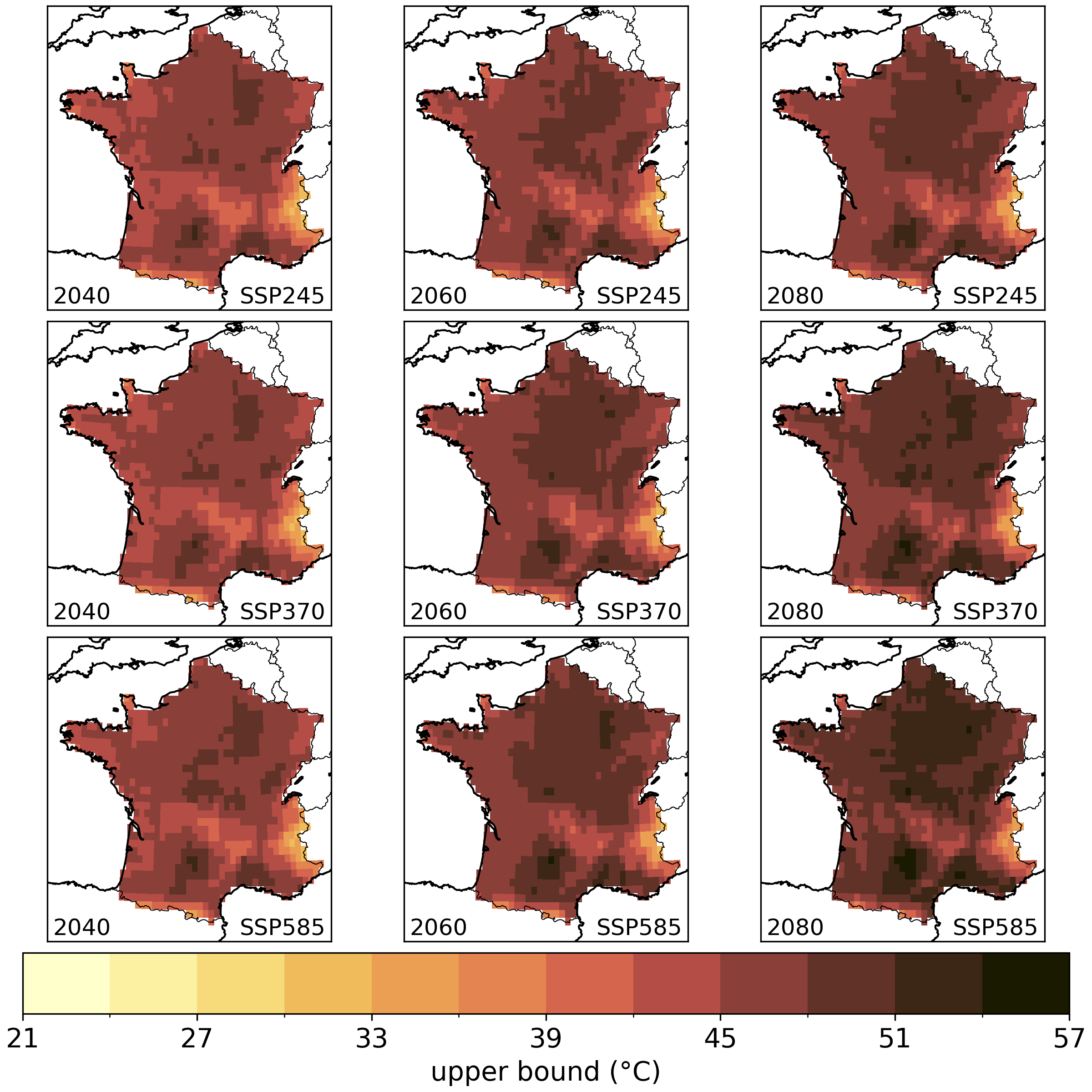}
	\caption{\label{fig:map:upper:bound}
		Maps of continental France for the upper bound
		(see Eq. \eqref{eq:gev:upper:bound}) of the GEV distribution for the annual 
		maxima of daily maximum surface air temperature values in the years 2040, 
		2060 and 2080 (columns) and for the climate scenarios SSP2-4.5, SSP3-7.0 and SSP5-8.5 (rows).
		Over all depicted maps, the values of the upper bound range 
		from 31.0 $^{\circ}$C to 56.7 $^{\circ}$C. The associated uncertainties (not shown here)
        range from 0.8 $^{\circ}$C to 8.0 $^{\circ}$C.}
\end{figure*}

\begin{figure*}[t]
    \centering
	\includegraphics[width=\linewidth]{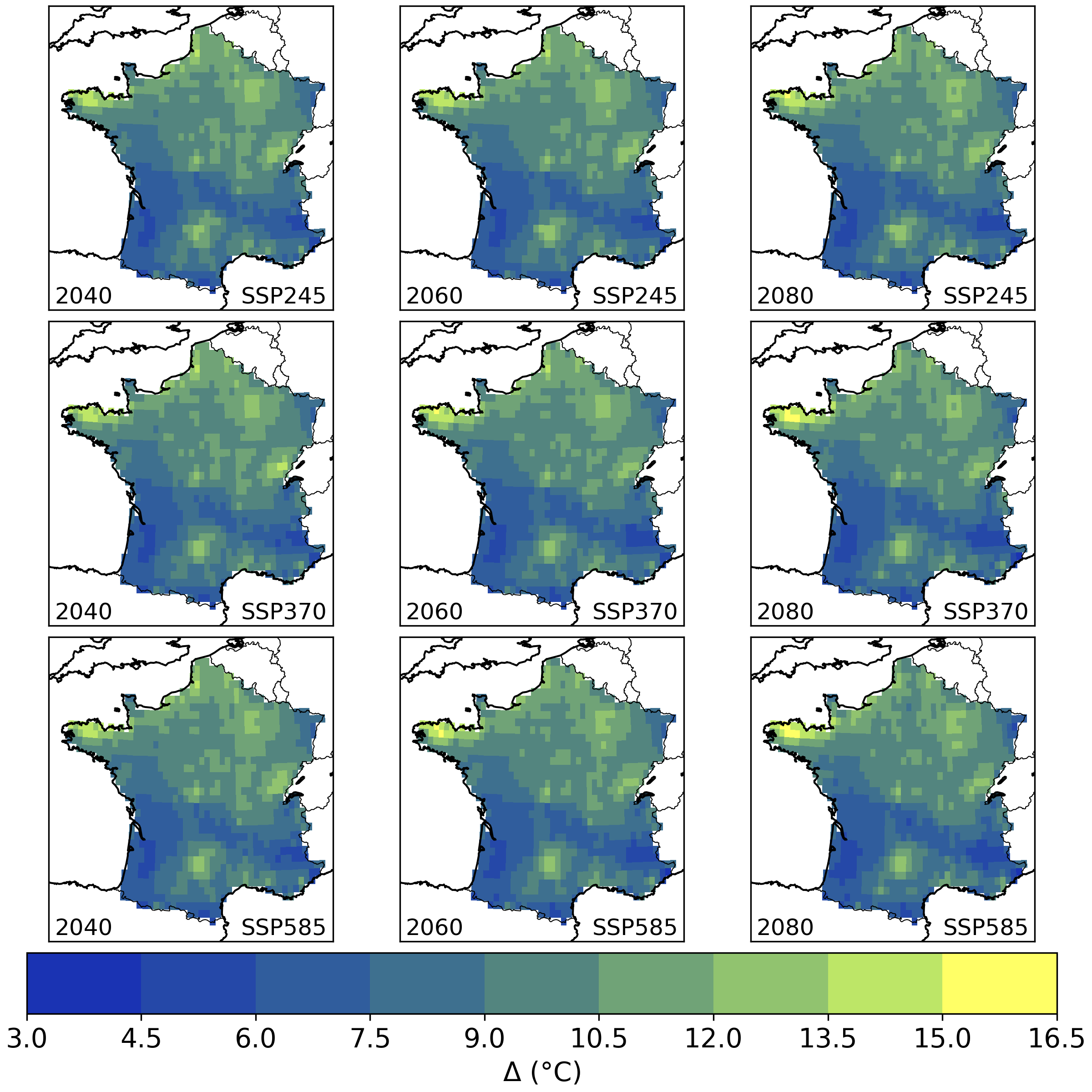}
	\caption{\label{fig:map:delta}
	Maps of continental France for the difference $\Delta_t$ (see Eq. 
	\eqref{eq:gev:delta}) between upper bound and median of the GEV 
	distribution for the annual maxima of daily maximum surface air temperature values in 
	the years 2040, 2060 and 2080 (columns) and for each 
	year with the climate scenarios SSP2-4.5, SSP3-7.0 and SSP5-8.5 (rows).
	Over all depicted maps, the values of $\Delta_t$ range 
	from 3.9 $^{\circ}$C to 16.3 $^{\circ}$C. The associated uncertainties (not shown here)
    range from 1.4 $^{\circ}$C to 9.8 $^{\circ}$C.}
\end{figure*}

For every single grid point, there is a corresponding non-stationary GEV distribution estimate, 
analogous to the one of Figure \ref{fig:explanatory:plot}.
The presented maps are constructed from these by evaluating the median or upper bound at the year in 
question, again at all grid points, and representing the obtained values as colors.\\
The median of the non-stationary GEV distribution of the annual maximum of daily maximum temperatures for the years 2040, 2060, and 2080 in continental France is depicted in Fig. 
\ref{fig:map:median}.
Depending on the specific location, annual maximum temperatures range from around 22 $^{\circ}$C to 45 $^{\circ}$C, a span of 23 $^{\circ}$C.
Overall, the south and south-west regions are estimated to experience hotter temperature extremes,
while the northern coast and mountainous regions, i.e, the Alps, are cooler in comparison.
These variations are expected due to natural differences in local climates and are not a specific feature of GEV distributions.\\
The upper bound and its difference $\Delta_t$ with respect to the median are shown
in Fig. \ref{fig:map:upper:bound} and Fig. \ref{fig:map:delta}, respectively.
Despite having rather large uncertainties, an observation for $\Delta_t$ is that the maps are qualitatively very similar, irrespective of year and climate 
scenario. This indicates that the GEV distribution over continental France is 
generally shifting towards higher temperatures while mostly maintaining its width.
Equivalent maps for the averaging period of $d=30$ days can be found
in the SM together with maps for annual minimum daily temperatures.\\
For the purpose of providing a more regional overview, Table 
\ref{tab:max:values:regions} lists the areal maximum values in the year 2080 for the administrative regions in continental France.
Note that the maximum for $\Delta_t$ is determined independently of the stated maximal median and maximal upper bound.
Occitanie (south-west of France) has the highest estimated GEV distribution median and upper bound over all investigated climate scenarios.
The coolest region in terms of high temperature extremes is Bretagne, with the overall lowest areal maximum median and upper bound.

\subsection{Underestimation of maximum high temperature extremes in TRACC}
The purpose of TRACC is to provide national, regional, and local authorities, infrastructure operators, and diverse stakeholders in France with a common reference for developing adaptation strategies for future climates.
The dataset on which TRACC is based consists of 17 simulations, 
combined from 9 regional and 6 global CMIP5 climate models, and is chronologically
close to an intermediate emission SSP2-4.5 scenario \citep{tracc_report_part1}.
Further details, especially on the differences to CMIP6, can be found in \citep{marson_2026}.
In TRACC, there are time horizons at which mainland France is projected
to reach a certain level of warming, i.e., an average temperature increase,
compared to the pre-industrial reference period (1850 to 1900).
These time horizons are: +2 $^{\circ}$C in 2030, + 2.7 $^{\circ}$C in 2050 and
+4 $^{\circ}$C in 2100, in line with a global warming level of 3°C by 2100.
For these years, the maximum and median over the entire TRACC dataset of the
annual daily maximum temperature, i.e., the hottest day of the year,
is shown in part two of the TRACC report \citep{tracc_report_part2}, namely in figure 12.
We compare our results to this report by visually inspecting the relevant years (2030, 2050, and 2100) in the TRACC map plots.
For this purpose, maps for continental France of the analysis results can be found in the SM with
the same colour selection to represent temperatures as in TRACC.
The median annual maximum temperatures are shown in SM Fig. 1, the 100-year return level
$\mathrm{Q}_t(0.01|\boldsymbol{\theta}_*)$ in SM Fig. 2, and the upper bound of annual maximum temperatures in SM Fig. 3.
Note that the year 2099 is used to compare with the data from TRACC in the year 2100.

For the SSP2-4.5 scenario, the median values are compatible within statistical error to the results in this paper,
except for the year 2099 where TRACC values are roughly +1$^{\circ}$C to +2$^{\circ}$C higher and more akin to the SSP3-7.0 scenario values.
The upper bound (see Eq. \eqref{eq:gev:upper:bound}) is used to compare with the annual daily maximum temperature in TRACC: for the years in question and the SSP2-4.5 scenario, the maps in the SM are mostly black since at almost all grid points the upper bound values are above 
the 44$^{\circ}$C threshold for colouring.
This indicates that the upper bound is either within a 2$^{\circ}$C range or exceeds the maximum temperature in TRACC almost everywhere.
In the year 2099, for example, the highest upper bound over all continental France for the SSP2-4.5 scenario was found to be 54(7)$^{\circ}$C and for the 
SSP3-7.0 scenario it is 57(8)$^{\circ}$C, both located in the Occitanie region.
These values exceed the maximum temperature stated in TRACC for this region, which is 46$^{\circ}$C, by either +8(7)$^{\circ}$C or +11(8)$^{\circ}$C.

For the TRACC maximum values, it is important to note that they reflect only the values actually encountered in the analysed data ensemble.
This is in contrast to the upper bound considered here, which is an extrapolated quantity
because it was determined through the GEV fitting procedure applied to the data.
The central statement of this section is therefore that the maximum temperature values reported in TRACC are misleading in the sense that they usually 
underestimate the statistically maximum possible temperature.
Or in other words, it is statistically possible to encounter higher temperature values than those stated in TRACC.
However, it should be noted that the worst-case conditions, i.e., temperatures close to the upper bound, have a very low probability and events with a 
lower return time for the year in question, e.g., 100 years, are usually more relevant.
A comparison of the 100-year return level with the TRACC maximum values shows that the 100-year return levels are around +2$^{\circ}$C warmer
for the SSP2-4.5 scenario and in all years, which is still within the read-off error from the colourbar.
Therefore, we conclude that the TRACC maximum temperature values are actually closer to events with a 100-year return time.

\subsection{Power infrastructure case studies}
\begin{sidewaysfigure*}[t]
	\includegraphics[width=\linewidth]{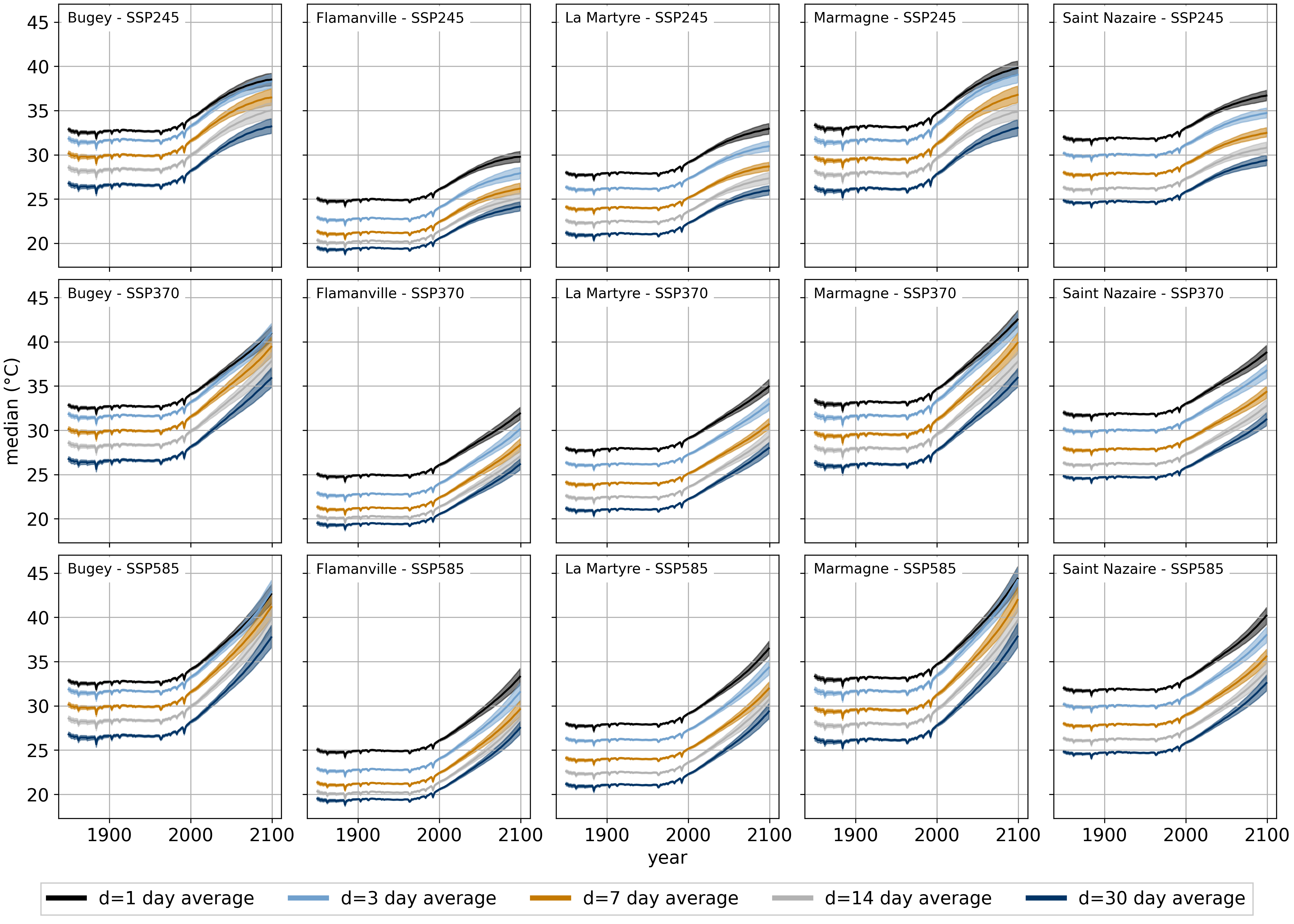}
	\caption{\label{fig:location:median}
	Median $\mathrm{Q}_t(0.5|\boldsymbol{\theta}_*)$ (see Eq. \eqref{eq:gev:quantile}) of the 
	non-stationary GEV distribution as a function of time for various averaging 
	periods $d$ at the infrastructure locations stated in Table 
	\ref{tab:infrastructure:locations}.
	The rows show the climate scenarios in order SSP2-4.5, SSP3-7.0 and 
	SSP5-8.5.
	The shaded areas represent the uncertainty of the estimate.}
\end{sidewaysfigure*}

\begin{sidewaysfigure*}[t]
	\includegraphics[width=\linewidth]{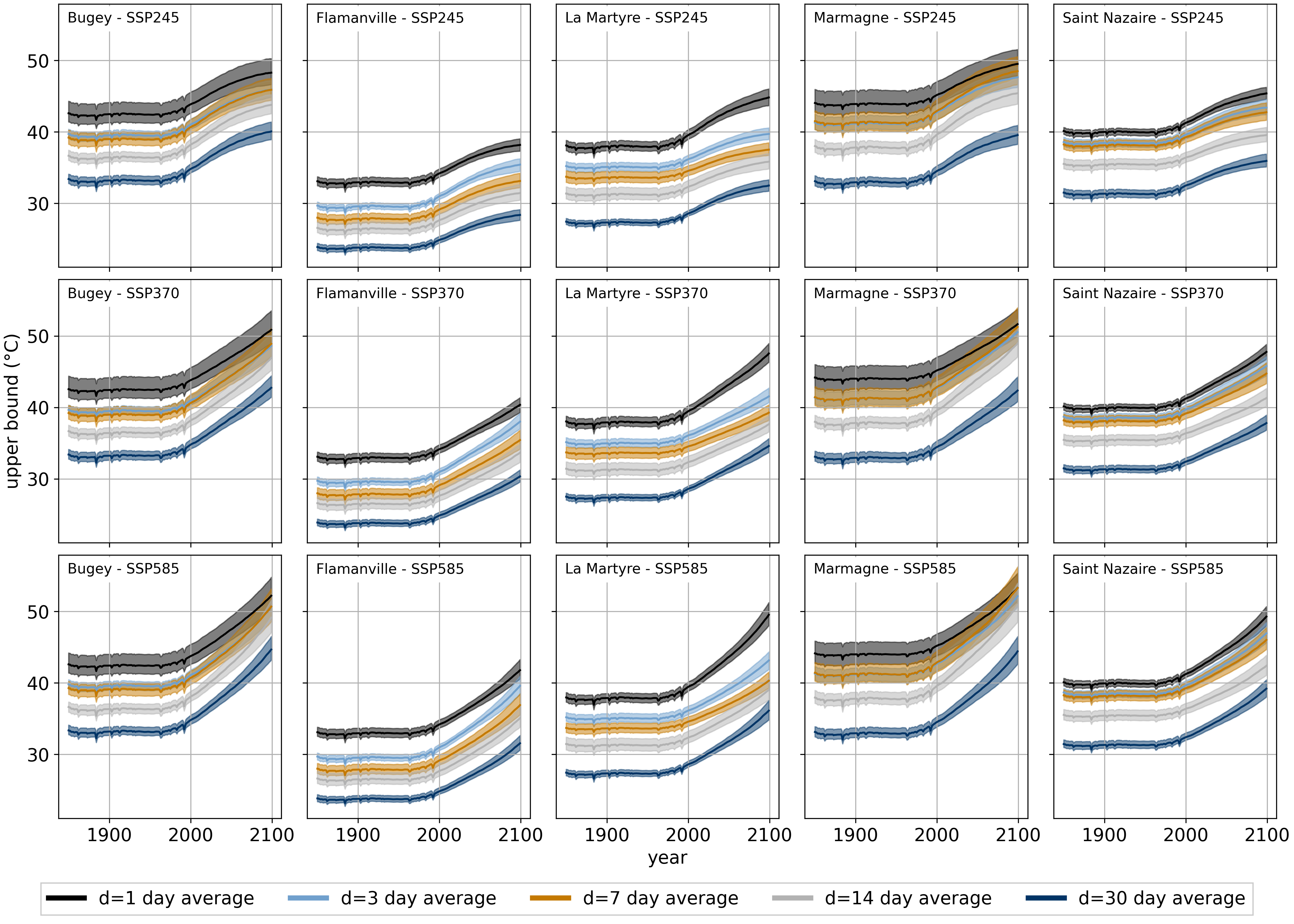}
	\caption{\label{fig:location:upper:bound}
	Upper bound (see Eq. \eqref{eq:gev:upper:bound}) of the non-stationary GEV 
	distribution as a function of time for various averaging periods $d$ at the 
	infrastructure locations stated in Table \ref{tab:infrastructure:locations}.
	The rows show the climate scenarios in order SSP2-4.5, SSP3-7.0 and 
	SSP5-8.5.
	The shaded areas represent the uncertainty of the estimate.}
\end{sidewaysfigure*}

We now investigate the time evolution of temperature maxima for a selection of electrical power infrastructure locations.
The corresponding coordinates for the locations are listed in Table \ref{tab:infrastructure:locations}.
For each location, the median and the upper bound are shown in Fig. \ref{fig:location:median} and
Fig. \ref{fig:location:upper:bound} respectively for all considered climate scenarios.
Each panel contains curves for averaging periods of $d=1, 3, 7, 14, 30$ days, which allow to assess 
high temperature events of longer duration.
Equivalent plots for the difference between upper bound and median $\Delta_t$ are available in the SM.
The curves are ordered from higher temperatures at short averaging periods to lower temperatures
at longer averaging periods for both the median and the upper bound.
The two exceptions to this rule are the upper bound at Bugey and Marmagne, where the ordering of the curve is less clear at the end of the century,
also because of the larger error bars.

\textit{Bugey \& Flamanville} are nuclear power plants, whose cooling systems must respect
certain security margins.
By 2099 and under the SSP5-8.5 scenario, the annual maximum of daily maximum temperatures is expected to exceed 
43$^{\circ}$C at Bugey and 34$^{\circ}$C at Flamanville with 50\% probability, while temperatures should 
statistically not exceed 51$^{\circ}$C at Bugey and 41$^{\circ}$C at Flamanville.\\
\textit{La Martyre} is the location where the Celtic link submarine power cable for the France-Ireland inter-connector is connected to the French network. 
Excess heat at this power station can damage some of its components and diminish the thermal rating of the 
cables, i.e., impose a reduction of the current that flows through the cables to prevent overheating.
At this location and for daily values ($d=1$), the median is expected to reach 36$^{\circ}$C and the 
upper bound 50$^{\circ}$C by 2099 for the SSP5-8.5 scenario.\\
\textit{Marmagne} is roughly situated in the center of France and the 400kV power line around this location
is undergoing an adaptation process to deal with the increase in newly installed capacities of renewable energy sources, the increase in trans-European energy flows and higher power consumption \citep{marmagne_project}.
As already mentioned in the introduction, since cables dilate under high temperatures, the cables themselves and the pylons, to which they are attached, must be designed such that a safety distance of the cables to the ground is maintained. To reduce the associated risk to people, the environment and livestock, the power flow through the cables can be reduced as a preventative measure in case of a high temperature event.
While a lower current would cause less heating of the cables through electric resistivity, it would also limit their energy transport capacity, i.e., the power line exploitation would be sub-optimal.
The median of the annual maximum of daily maximum temperature is projected to reach 44$^{\circ}$C and the estimated upper bound is around 52$^{\circ}$C by the year 2099 for the SSP5-8.5 scenario.
It should also be noted that overhead lines are usually located approximately 30 to 90 m above the ground, higher than the 2 m temperature used in this study.
Solar irradiance is another important parameter that influences cables' temperature.\\
\textit{Saint Nazaire} is the connection point of the offshore wind farm to the inland power network.
Regarding high temperatures, it is subject to similar constraints as La Martyre.
The median in the year 2099 is 40$^{\circ}$C for the SSP5-8.5 scenario and averaging period $d=1$. The corresponding upper bound is close to 50$^{\circ}$C.

\section{Conclusions}
In this paper, we demonstrated the effectiveness of the Bayesian approach to non-stationary GEV parameter estimation in \citep{robin_2026} for risk assessment of electrical power infrastructure.
Specifically, the statistics of temperature extremes were estimated, including error bars for three climate scenarios, up to the year 2099 and with 0.25 degree spatial resolution over continental France.
The method allows combining simulation data from 28 CMIP6 climate models and E-OBS observational records into a single holistic estimate.

We portrayed the spatial structure of temperature extremes over continental France through maps of the median and the upper bound of the GEV distribution, while additional summary information was provided for each administrative region in France.
The general observation was that the south-west of France is subject to higher temperature extremes than the rest of the country, while costal and mountainous regions usually have less intense annual maximum temperature values.
For five exemplary power infrastructure locations, we investigated the temporal evolution of temperature extremes highlighting the usefulness of the approach to give estimates and uncertainties at arbitrary locations and for a time horizon that potentially spans the lifetime of the asset/structure in question.
Furthermore, a comparison of the results of this paper to those of TRACC was performed, giving an indication that the maximum temperature values stated in TRACC could actually be exceeded in the future.

An interesting future application of the method would be to consider extreme wind speeds.
Statistics on high wind speeds are relevant, since strong winds can damage wind turbines, overhead lines, and pylons.
Also, both too high and too low wind speeds can affect energy production: at too high wind speeds, wind turbines are turned off to prevent potential damage, and at too low wind speeds, they stop rotating.
Another perspective could be the application of the method directly to the balance of power supply and demand in the electrical energy system.
This could give statistical information on peak loads, including their uncertainties.
Finally, extremes can also be estimated using climate emulators and rare events sampling methodologies, as in \citep{lancelin_2026}. These methods allow fast computing of very rare events, and present the advantage to have lower uncertainty ranges.

\section*{CRediT authorship contribution statement}
\textbf{Peter Werner:} Conceptualization, Visualization, Investigation, Writing - Original Draft, Writing - Review \& Editing, Software, Formal analysis.
\textbf{Yoann Robin:} Conceptualization, Writing - Review \& Editing, Methodology, Software, Validation.
\textbf{Laurent Dubus:} Conceptualization, Writing - Review \& Editing, Validation, Supervision, Funding acquisition.
\textbf{Freddy Bouchet:} Conceptualization, Writing - Review \& Editing, Validation, Supervision, Funding acquisition.
\textbf{Bastien Cozian:} Writing - Review \& Editing, Validation.

\section*{Funding}
This work has been funded by the research collaboration contract C22/1764 between RTE and
IPSL/Sorbonne Université.

\section*{Declaration of competing interest}
The authors declare no conflicts of interest relevant to this study.

\section*{Acknowledgements}
To process the data, this study benefited from the IPSL Data and Computing 
Center ESPRI, which is supported by CNRS, SU, CNES and Ecole Polytechnique.\\
The authors acknowledge the IPSL computing and data center ESPRI for their 
support in supplying and accessing the data.\\
This work was partially performed using HPC resources from GENCI-IDRIS
(Grant 2025-104472), too.\\
The authors acknowledge the E-OBS dataset from the EU-FP6 project UERRA 
(https://www.uerra.eu) and the Copernicus Climate Change Service, and the data 
providers in the ECA\&D project(https://www.ecad.eu).
\clearpage

\clearpage

\clearpage

\clearpage

\clearpage

\clearpage

\clearpage

\clearpage

\begin{sidewaystable*}[t]
	\caption{\label{tab:max:values:regions}
	Maximum value of the median, upper bound and difference $\Delta_t$ (see Eq. \eqref{eq:gev:delta}) in 
	the year 2080 over every administrative region in continental France for the
	climate scenarios SSP2-4.5, SSP3-7.0 and SSP5-8.5. Note that the
	maximum for $\Delta_t$ is determined independently from the stated maximal median and maximal upper bound. Uncertainties, i.e., the inter quartile range, are given within the trailing parenthesis.}
    \small
	\begin{tabular}{l|ccc|ccc|ccc}
	\hline
	 & \multicolumn{3}{c|}{SSP2-4.5 [$^{\circ}$C]} & \multicolumn{3}{c|}{SSP3-7.0 [$^{\circ}$C]} & \multicolumn{3}{c}{SSP5-8.5 [$^{\circ}$C]}\\
	Region & Median & Upper bound & $\Delta_t$ & Median & Upper bound & $\Delta_t$ & Median & Upper bound & $\Delta_t$\\
	\hline
	Auvergne-Rhône-Alpes & 41(1) & 51(5) & 11(5) & 43(1) & 52(4) & 11(7) & 44(2) & 53(5) & 11(6) \\
	Bourgogne-Franche-Comté & 40(1) & 51(5) & 13(7) & 41(2) & 52(5) & 13(7) & 42(2) & 53(5) & 13(7) \\
	Bretagne & 37(1) & 48(3) & 15(4) & 39(1) & 50(4) & 16(6) & 40(1) & 52(5) & 16(6) \\
	Centre-Val de Loire & 40(1) & 51(6) & 12(7) & 41(2) & 53(6) & 12(8) & 43(2) & 55(6) & 12(8) \\
	Grand Est & 39(1) & 51(5) & 13(7) & 41(2) & 53(5) & 13(7) & 42(2) & 54(6) & 13(7) \\
	Hauts-de-France & 39(1) & 50(5) & 13(6) & 40(2) & 51(5) & 13(6) & 42(2) & 53(5) & 13(7) \\
	Île-de-France & 39(1) & 50(3) & 11(5) & 41(1) & 51(4) & 11(5) & 42(2) & 52(4) & 11(6) \\
	Normandie & 38(1) & 49(3) & 13(5) & 40(2) & 51(3) & 14(5) & 41(2) & 52(4) & 14(6) \\
	Nouvelle-Aquitaine & 41(2) & 50(5) & 11(7) & 42(1) & 51(5) & 11(7) & 43(2) & 52(6) & 11(7) \\
	Occitanie & 42(2) & 54(7) & 13(8) & 43(1) & 55(7) & 13(9) & 45(2) & 57(8) & 13(10) \\
	Pays de la Loire & 39(1) & 49(4) & 10(5) & 41(1) & 50(4) & 11(5) & 41(1) & 51(4) & 11(6) \\
	Provence-Alpes-Côte d'Azur & 41(1) & 51(4) & 12(6) & 43(1) & 52(4) & 12(7) & 45(2) & 53(3) & 12(8) \\
	\hline
	\end{tabular}
\end{sidewaystable*}
\clearpage

\begin{table}[t]
    \centering
	\caption{\label{tab:infrastructure:locations}
	Investigated infrastructure locations in France. Given are the latitudinal 
	and longitudinal coordinates and a short description of the facility.}
    \footnotesize
	\begin{tabular}{l|c|c|c}
		\hline
		Name & Latitude & Longitude & Description \\
		\hline
		Bugey & 45.80$^{\circ}$ N & 5.27$^{\circ}$ E & nuclear power plant\\
		Flamanville & 49.54$^{\circ}$  N & 1.88$^{\circ}$ W & nuclear power plant\\
		La Martyre & 48.44$^{\circ}$ N & 4.20$^{\circ}$  W & power station at Celtic link inter-connector\\
		Marmagne & 47.10$^{\circ}$ N & 2.26$^{\circ}$  E & power station\\
		Saint Nazaire & 47.36$^{\circ}$  N & 2.02$^{\circ}$  W & connection station to Saint Nazaire offshore wind farm\\
		\hline
	\end{tabular}
\end{table}
\clearpage

\appendix
\section{Generalized extreme value distribution} 
\label{sec:gev:distribution}
The probability density function of the GEV distribution is given by
\begin{align}
	\textrm{GEV}(T|\mu, \sigma, \xi) &= \frac{1}{\sigma} {\tau(T)}^{\xi+1} \exp{\left(-\tau(T)\right)}\\
	\tau(T) &= \left\lbrace \begin{array}{cl}
		\left[ 1 + \xi \left( \frac{T - \mu}{\sigma}\right) \right]^{-1/\xi} & \text{if } \xi \neq 0\\
		\exp{\left(-\frac{T-\mu}{\sigma}\right) } & \text{if } \xi = 0 \nonumber
	\end{array}\right. ,
\end{align}
with location $\mu \in \mathbb{R}$, scale $\sigma > 0$, and shape $\xi \in \mathbb{R}$ parameters.
The random variable $T$ can take values over the following support ranges:
\begin{equation}
	T \in \left\lbrace
	\begin{array}{cl}
		[ \mu - \frac{\sigma}{\xi}, +\infty ) & \text{for }
		\xi > 0 \\
		( -\infty, +\infty ) & \text{for } \xi = 0 \\
	 	( -\infty, \mu - \frac{\sigma}{\xi} ] &
		\text{for } \xi < 0
	\end{array}
	\right. .
\end{equation}
Note that according to the context, the expression $\mu - \sigma / \xi$ is referred
to as either the \emph{upper bound} or \emph{lower bound}.

\section{List of used CMIP6 models}
\linespread{1.0}
\begin{table*}[t]
\centering
\caption{\label{tab:cmip6:models} List of the 28 CMIP6 models used 
	to determine the a priori distribution necessary for the Bayesian GEV 
	distribution parameter estimation procedure.}
    \scriptsize
    \begin{tabular}{p{0.7\textwidth}%
		>{\raggedleft\arraybackslash}p{0.13\textwidth}%
		>{\raggedleft\arraybackslash}p{0.14\textwidth}}
    \hline
	Modeling Center or Group & Institute ID & Model name\\
	\hline
	
	Research Center for Environmental Changes, Academia Sinica,
	Nankang, Taipei 11529, Taiwan & AS-RCEC & TaiESM1 \\
	
	Alfred Wegener Institute, Helmholtz Centre for Polar and Marine 
	Research, Am Handelshafen 12, 27570 Bremerhaven, 
	Germany & AWI & AWI-CM-1-1-MR \\
	
	Beijing Climate Center, Beijing 100081, China & BCC & 
	BCC-CSM2-MR \\
	
	Chinese Academy of Meteorological Sciences, Beijing 100081, 
	China & CAMS & CAMS-CSM1-0 \\
	
	Chinese Academy of Sciences, Beijing 100029, China & CAS & 
	FGOALS-g3 \\
	
	Canadian Centre for Climate Modelling and Analysis, Environment 
	and Climate Change Canada, Victoria, BC V8P 5C2, Canada & CCCma 
	& CanESM5 \\
	
	Fondazione Centro Euro-Mediterraneo sui Cambiamenti Climatici, 
	Lecce 73100, Italy & CMCC & CMCC-ESM2 \\
	
	CNRM (Centre National de Recherches 
	Meteorologiques, Toulouse 31057, France), CERFACS (Centre 
	Europeen de Recherche et de Formation Avancee en Calcul 
	Scientifique, Toulouse 31057, France)
	& CNRM-CERFACS & CNRM-CM6-1 CNRM-ESM2-1 \\
	
	Commonwealth Scientific and Industrial Research Organisation, 
	Aspendale, Victoria 3195, Australia & CSIRO  & ACCESS-ESM1-5 \\
	
	CSIRO (Commonwealth Scientific and Industrial Research 
	Organisation, Aspendale, Victoria 3195, Australia), ARCCSS 
	(Australian Research Council Centre of Excellence for Climate 
	System Science) & CSIRO-ARCCSS  & ACCESS-CM2 \\
	
	AEMET, Spain; BSC, Spain; CNR-ISAC, Italy; DMI, 
	Denmark; ENEA, Italy; FMI, Finland; Geomar, Germany; ICHEC, 
	Ireland; ICTP, Italy; IDL, Portugal; IMAU, The Netherlands; 
	IPMA, Portugal; KIT, Karlsruhe, Germany; KNMI, The 
	Netherlands; Lund University, Sweden; Met Eireann, Ireland; 
	NLeSC, The Netherlands; NTNU, Norway; Oxford University, 
	UK; surfSARA, The Netherlands; SMHI, Sweden; Stockholm 
	University, Sweden; Unite ASTR, Belgium; University College 
	Dublin, Ireland; University of Bergen, Norway; University 
	of Copenhagen, Denmark; University of Helsinki, Finland; 
	University of Santiago de Compostela, Spain; Uppsala 
	University, Sweden; Utrecht University, The Netherlands; 
	Vrije Universiteit Amsterdam, the Netherlands; Wageningen 
	University, The Netherlands. Mailing address: EC-Earth 
	consortium, Rossby Center, Swedish Meteorological and 
	Hydrological Institute/SMHI, SE-601 76 Norrkoping, Sweden 
	& EC-Earth-Consortium & EC-Earth3 EC-Earth3-Veg-LR 
	EC-Earth3-Veg\\
	
	Institute for Numerical Mathematics, Russian 
	Academy of Science, Moscow 119991, Russia
	& INM & INM-CM4-8 INM-CM5-0 \\
	
	Institut Pierre Simon Laplace, Paris 75252, France & IPSL & 
	IPSL-CM6A-LR \\
	
	JAMSTEC (Japan Agency for Marine-Earth Science 
	and Technology, Kanagawa 236-0001, Japan), AORI (Atmosphere 
	and Ocean Research Institute, The University of Tokyo, 
	Chiba 277-8564, Japan), NIES (National Institute for 
	Environmental Studies, Ibaraki 305-8506, Japan), and R-CCS 
	(RIKEN Center for Computational Science, Hyogo 650-0047, 
	Japan) & MIROC & MIROC6 MIROC-ES2L \\
	
	Met Office Hadley Centre, Fitzroy Road, Exeter, Devon, EX1 
	3PB,UK & MOHC & UKESM1-0-LL \\
	
	Max Planck Institute for Meteorology, Hamburg 20146, Germany 
	&MPI-M & MPI-ESM1-2-LR \\
	
	Meteorological Research Institute, Tsukuba, Ibaraki 305-0052, 
	Japan & MRI & MRI-ESM2-0 \\
	
	Goddard Institute for Space Studies, New York, NY 10025, USA & 
	NASA-GISS & GISS-E2-1-G \\
	
	NorESM Climate modeling Consortium consisting of CICERO (Center 
	for International Climate and Environmental Research, Oslo 
	0349), MET-Norway (Norwegian Meteorological Institute, Oslo 
	0313), NERSC (Nansen Environmental and Remote Sensing Center, 
	Bergen 5006), NILU (Norwegian Institute for Air Research, 
	Kjeller 2027), UiB (University of Bergen, Bergen 5007), UiO 
	(University of Oslo, Oslo 0313) and UNI (Uni Research, Bergen 
	5008), Norway. Mailing address: NCC, c/o MET-Norway, Henrik 
	Mohns plass 1, Oslo 0313, Norway & NCC & 
	NorESM2-LM  NorESM2-MM \\
	
	National Institute of Meteorological Sciences/Korea 
	Meteorological Administration, Climate Research Division, 
	Seoho-bukro 33, Seogwipo-si, Jejudo 63568, Republic of Korea & 
	NIMS-KMA & KACE-1-0-G UKESM1-0-LL\\
	
	National Oceanic and Atmospheric Administration, Geophysical 
	Fluid Dynamics Laboratory, Princeton, NJ 08540, USA & 
	NOAA-GFDL  & GFDL-ESM4 \\
\hline
\end{tabular}
\end{table*}
\linespread{1.5}
\clearpage

\bibliographystyle{elsarticle-harv}
\bibliography{references.bib}
\clearpage

\end{document}